\documentclass[fleqn,10pt]{wlscirep}
\usepackage[utf8]{inputenc}
\usepackage[T1]{fontenc}
\usepackage{bm}
\usepackage{ dsfont }

\title{Synthetic dimensions for topological and quantum phases: Perspective}

\author[1]{Javier Argüello-Luengo}
\author[1]{Utso Bhattacharya}
\author[2]{Alessio Celi}
\author[3]{Ravindra W. Chhajlany}
 \author[4,5,1]{Tobias Grass}
\author[1]{Marcin P\l odzie\'n}
\author[6]{Debraj Rakshit}
\author[1]{Tymoteusz Salamon}
\author[1]{Paolo Stornati}
\author[1,7]{Leticia Tarruell}
\author[1,7,*]{Maciej Lewenstein}

\affil[1]{ICFO - Institut de Ciencies Fotoniques, The Barcelona Institute of Science and Technology, Avenida Carl Friedrich Gauss 3, E-08860 Castelldefels (Barcelona), Spain}
\affil[2]{Departament de Física, Universitat Autónoma de Barcelona, E-08193 Bellaterra, Spain}
\affil[3]{Institute of Spintronics and Quantum Information, Faculty of Physics, Adam Mickiewicz University,  61-614 Pozna{\'n}, Poland }
\affil[4]{DIPC - Donostia International Physics Center, Paseo Manuel de Lardizabal 4, 20018 San Sebastian, Spain }
\affil[5]{Ikerbasque - Basque Foundation for Science, Maria Diaz de Haro 3, 48013 Bilbao, Spain}
\affil[6]{Harish-Chandra Research Institute, A CI of Homi Bhabha National Institute, Chhatnag Road, Jhunsi, Allahabad 211019, India}
\affil[7]{ICREA, Passeig Lluis Companys 23, 08010 Barcelona, Spain}
\affil[*]{e-mail: maciej.lewenstein@icfo.eu}

\begin{abstract}
In this Perspective article we report on recent progress on studies of synthetic dimensions, mostly, but not only,  based on the research realized around the Barcelona groups (ICFO, UAB), Donostia (DIPC), Pozna\'n (UAM), Kraków (UJ), and Allahabad (HRI). The concept of synthetic dimensions works particularly well in atomic physics, quantum optics, and photonics, where the internal degrees of freedom (Zeeman sublevels of the ground state, metastable excited states, or motional states for atoms, and angular momentum states or transverse modes for photons) provide the synthetic space. We describe our attempts to design quantum simulators with synthetic dimensions, to mimic curved spaces, artificial gauge fields, lattice gauge theories, twistronics, quantum random walks, and more.
\end{abstract}
\begin{document}

\flushbottom
\maketitle
 
\thispagestyle{empty}

\noindent \textbf{Key points:}

\begin{itemize}
    \item Quantum simulators employing synthetic dimensions pave the way for mimicking the exotic space-times.

    \item Synthetic dimensions in ultra-cold quantum gases in optical lattices allow to study of artificial gauge fields and lattice gauge theories.

    \item Synthetic dimensions in ultra-cold quantum gases in optical lattices permit to simulate ``twistronics without a twist''

    \item Quantum random walks allow the study of a broad spectrum of symmetry-protected topological phases observed in photonic systems in one or two spatial dimensions.

    \item Discrete time crystals are time-periodic phases of matter realizing a new platform for quantum simulators, where time can be used as an additional artificial dimension, allowing studies of high dimensional topological models.

    \item In recent years enormous experimental progress in realizing synthetic dimensions took place.

\end{itemize}

\noindent \textbf{Website summary:} 

Quantum simulators study important models of condensed matter and high-energy physics. Research on synthetic dimensions has paved the way for studying exotic phenomena, such as curved space-times, topological phases of matter, lattice gauge theories, twistronics without a twist, and more.

\section{Introduction \label{Sec1}}
The use of internal atomic states as an effective synthetic dimension is an idea introduced in 2011\cite{Boada_2011} that gained quite a popularity and maturity in the last years. This is one of the main motivation of this Perspective article, which obviously has elements of review, but mostly cover the material related to the works of the Quantum Optics Theory and Quantum Gas Experiments groups at ICFO,  but also other groups in Barcelona (UAB), Donostia (DIPC), Pozna\'n (UAM), Kraków (UJ), and Allahabad (HRI). Synthetic dimensions have already been reported in several reviews, such as the recent Quick Study in Physics Today by K. Hazzard and B. Gadway \cite{Hazzard2023}, who write: \emph{``Frequently, synthetic dimensions are created in ultrasmall and ultracold systems, where the experiments provide powerful access to the hard-to-understand world of interacting quantum matter, which underpins fields as diverse as quantum gravity, solid-state physics... Objects move through three dimensions in space. But a wide range of experiments that manipulate atoms, molecules, and light can engineer artificial matter in ways that break even that basic law of nature. Such matter can behave as if it were extended to four or more spatial dimensions or restricted to just one or two, as determined by experimental design.''}

Boada {\it et al.} made a ``trivial'', yet very deep observation that dimension of a lattice depends on its connectivity \cite{Boada_2011}. This allowed them to mimic 4D physics in 3D lattice. In the same paper, they already suggested that the internal states of particles involved in the dynamic on a D-dimensional lattice may be used to increase the dimension. This idea was fully developed in the seminal Letter \cite{PhysRevLett.112.043001}, where it was shown that the basic idea of synthetic dimensions naturally allows to introduction of artificial gauge field, corresponding to complex phase factors on the synthetic bonds (see Ref\cite{Price2019} for a review ). Synthetic dimensions can be realized in various platforms, from cold atoms in optical lattices \cite{Boada_2011,PhysRevLett.108.133001} through photonic systems \cite{Yuan2018}, Rydberg atoms \cite{Kanungu2022}, and more. 

This Perspective article is organized as follows: After a short  ``Introduction'', Section~\ref{Sec2} focuses on ``Original motivation: Quantum simulations of artificial space-times''. Section~\ref{Sec3} deals with ``Synthetic gauge field in synthetic dimension'', while  Section~\ref{Sec4} with ``Quantum simulators of Lattice Gauge Theories''. In Section~\ref{Sec5}, we discuss the use of synthetic dimensions for ``Twistronics''. Sections~\ref{Sec6} and \ref{Sec7} cover more ``exotic'' approaches to synthetic dimensions: those based on ``Quantum Random Walks'', and ``Time Crystal Platform for Quantum Simulations''. In Section~\ref{Sec8}, we discuss the experimental perspective on quantum simulators utilizing synthetic dimensions. We conclude in Section~\ref{Sec9}.

\section{Original motivation: Quantum simulations of artificial spacetimes \label{Sec2}}
Ultracold atom in optical lattices are marvelous quantum simulators of condensed matter models in almost arbitrary 1D, 2D, and 3D geometries.
However, optical lattices are generically straight and with open boundary conditions. Similar limitations apply to photonic simulators in waveguides and resonator arrays. In quantum field theory, we are also interested in curved spacetimes, spacetimes of higher dimensions, and of non-trivial topology to implement generalized boundary conditions. Thus, how to meet the quest for artificial spacetimes? 

In fact, for realizing lattice models living in curved spacetime, we do not need to bend the lattice. It suffices to spatially modulate the tunneling\cite{Boada_2011}:
in a model realizing Dirac fermions it causes a position-dependent Fermi velocity that corresponds to the motion in a curved spacetime called {\it optical}. Such a family of spacetimes includes the one seen by an accelerated observer described by the Rindler metric. Dirac fermions in the positive (negative) wedge of the Rindler metric have a Fermi velocity that grows (decreases) linearly in the direction of the acceleration. The surface orthogonal to the acceleration with zero Fermi velocity is the event horizon that separates the two wedges that are casually disconnected. A unique feature of artificially realized spacetimes is that they allow for quantum quenches of spacetime itself. By a sudden change from the ordinary Minkowski metric seen by an observer at rest in flat spacetime to the Rindler metric of an accelerated observer, one can simulate the celebrated Unruh effect for free\cite{PhysRevA.95.013627} and interacting\cite{10.21468/SciPostPhys.5.6.061} fermions: the vacuum (Dirac sea) appears to the accelerated observer as a thermal state, with a temperature proportional to the proper acceleration, that is, inversely proportional to the distance from the horizon and to the local Fermi velocity. In fact, as originally noted by Takagi, in two spatial dimensions, one can observe an apparent {\it inversion of statistics},  with thermal excitations following a Bose-Einstein distribution.

Similarly, we can overcome the limitations on the dimensionality and the boundary conditions by considering the coherent couplings between properly chosen internal degrees of freedoms. Such couplings can not only provide additional neighbor where tunneling, thus provide extra dimensions\cite{PhysRevLett.108.133001}. By removing the identification between tunneling and spatial displacement they allow for lattices with periodic and twisted boundary conditions
\cite{Boada_2015}, in a word for artificial spacetimes of non-trivial topology. This capability, together with the ability of suddenly quenching the geometry from the one of the torus to a Klein bottle, for instance, by a sudden change of the internal states' couplings, opens the possibility to statically and dynamically probing the effect of spacetime topology on many-body quantum phases.

\section{Synthetic gauge field in synthetic dimension\label{Sec3}}

\begin{figure}
\centering
    \includegraphics[width=0.9\columnwidth]{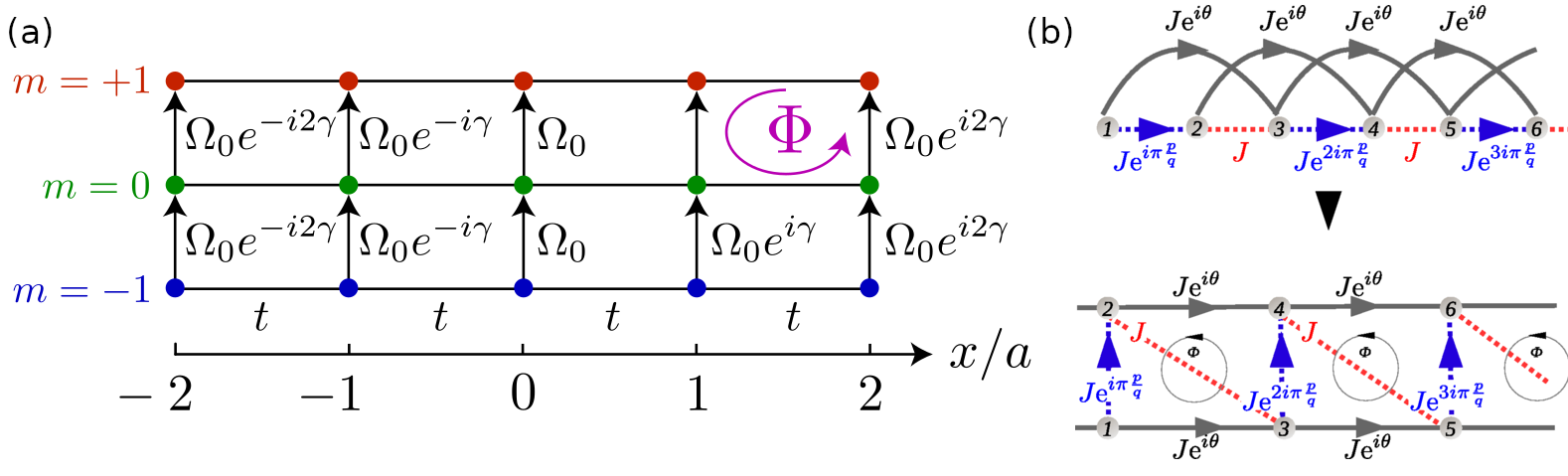}
    \caption{(a) In a optical 1D chain lattice, three atomic hyperfine levels are coupled through Raman beams with Rabi frequency $\Omega_0$ and spatially varying phase $x\gamma$. This artificial hopping in the synthetic dimension together with the real-space hopping $t$ mimics the Hamiltonian of a compact 2D square lattice with synthetic magnetic fluxes, in close analogy to the Hofstadter model. Figure taken from Ref.~\citeonline{PhysRevLett.112.043001}. (b) A 1D system with next-nearest-neighbor hopping $J$ can be mapped onto a two-leg ladder. Spatially varying hopping phases (in blue), which can be implemented through Floquet engineering, generate synthetic magnetic fluxes. The additional phase $\theta$ provides a cyclic Hamiltonian parameter which can be used to substitute momentum along a second dimension. Figure taken from Ref.~\citeonline{Grass_2015}. \label{fig:synsyn1}}
\end{figure}

The idea behind synthetic dimensions is to use an internal degree of freedom of a system, e.g. the electronic level of atoms, in order to mimic an additional external degree of freedom\cite{PhysRevLett.108.133001}. With this degree of freedom typically being a discrete one, the natural setting of the synthetic dimension is within a lattice system. The kinetic term in a lattice are hopping processes, often limited to nearest-neighbor processes, and hence the kinetic term in the synthetic dimension can be achieved by an optical coupling between the adjacent energy levels.
Quite naturally, the optical coupling comes along with a tunable space dependency, e.g. a position-dependent phase factor. For the hopping along the synthetic dimension this translates into a Peierls phase, e.g. a synthetic magnetic flux. In the original proposal, Ref.~\citeonline{PhysRevLett.112.043001}, it has been suggested to build in this way a Hofstadter-like model out of a 1D chain using the three hyperfine 
ground states of $^{87}$Rb for a compact second dimension, see Fig.~\ref{fig:synsyn1}(a). The compactness of the dimension produces sharp edges along the synthetic dimension which facilitate the detection of chiral edge states, one of the characteristic features of such a model. The proposal has then been realized in Ref.~\citeonline{Stuhl_2015} and Ref.~\citeonline{Mancini_2015}, the latter one using the $I=5/2$ nuclear spin manifold of fermionic $^{173}$Yb, thus allowing for up to a six-leg ladder in the synthetic dimension. 

The analogy between the kinetic term in a synthetic dimension and an optical coupling of internal levels is, to some extent, spoiled in interacting systems. For instance, contact interactions in real space may translate into ``long-range'' interactions within the synthetic dimension, that is, interactions without any spatial decay. There are ways of overcoming this limitation, although they require relatively complicated engineering of interactions \cite{Barbiero_2020}. However, the special structure of interactions might also be welcomed as a feature of the synthetic dimension.
As an example, we refer to the case of a synthetic bilayer, obtained from the optical coupling of two Landau levels in graphene \cite{Ghazaryan_2017}, which differs from real bilayer graphene by a modified interaction potential. It has been shown in Ref.~\citeonline{Cian_2020} that, at filling 2/3, the modified pseudopotentials support non-Abelian phases with Fibonacci anyons, rather than an Abelian Laughlin-state which would be supported by the single-layer system.

The concept of synthetic dimensions\cite{PhysRevLett.108.133001,PhysRevLett.112.043001,PhysRevA.90.043628,10.21468/SciPostPhys.15.2.046,PhysRevLett.128.070603} is very general, and internal atomic or electronic states are not the only candidates for simulating synthetic spatial degrees of freedom. For instance, in long-ranged systems with connectivity between spatially separated constituents, the connectivity graph can be viewed as a nearest-neighbor model embedded in some higher dimension. As a simplest example, consider a spin chain with nearest- and next-nearest-neighbor interactions. By interpreting even and odd sites of the chain as two legs of a two-leg-ladder, the chain is mapped onto a ladder, e.g. a structure which is beyond 1D, with couplings only between nearest neighbors. One can then apply Floquet engineering methods to compensate for the potentially undesired spatial decay of longer-range interactions. By appropriately adjusting the Floquet engineering protocol, one can also thread synthetic fluxes through the synthetic plaquettes of the ladder \cite{Grass_2015,Major2017,Grass_2018} by equipping some couplings with spatially dependent phase factors, see Fig.~\ref{fig:synsyn1}(b). Suitable systems to realize such a scheme are, for instance, trapped ions which are often limited to 1D, but offer long-range couplings and individual addressability to realize the described scheme. The synthetic flux provides a tool to realize synthetic topology\cite{Mochol-Grzelak_2019,10.21468/SciPostPhys.3.2.012}: to study fractal energy spectra similar to the Hofstadter butterfly and topological features such as chiral states and Chern numbers. In this context, a cyclic Hamiltonian parameter such as constant hopping phase $\theta$ [see Fig.~\ref{fig:synsyn1}(b)] might provide an analog for the momentum in the second dimension.

\section{Quantum simulators of Lattice Gauge Theories\label{Sec4}}

\begin{figure}
\centering
\includegraphics[width=0.9\columnwidth]{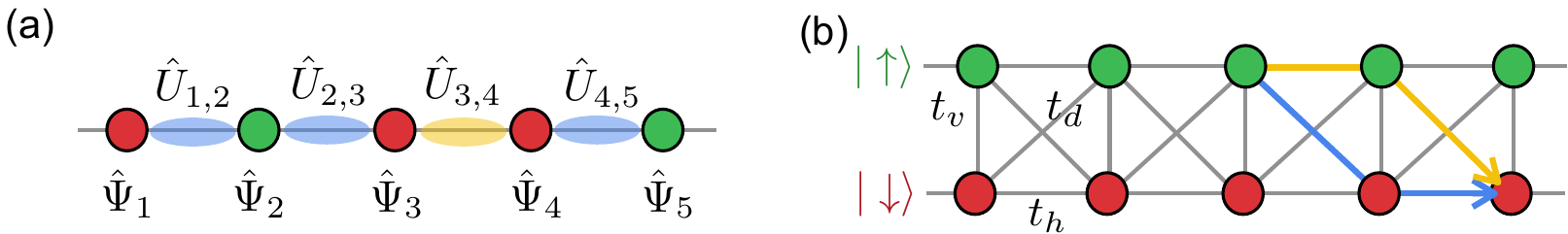}
    \caption{(a)  The Schwinger model in 1+1 dimension. The lattice comprises vertices containing matter particles (red and green) connected by links carrying the associated gauge field (blue for positive and yellow for negative values).  
    The Fermionic particles $\hat{\Psi}_i$ that can hope from one site to the other, with a hopping strength mediated by the gauge field $\hat{U}_{i,i+1}$, with an energy contribution in the Hamiltonian given by $\hat{\Psi}_i\hat{U}_{i,i+1}\hat{\Psi}^\dagger_{i+1}$ . The energy density of the electromagnetic field is given by the square of the Electric field. 
    (b) Two-leg ladder associated with the Creutz-ladder model. Without vertical tunneling ($t_v$), the two-site paths depicted in yellow and blue can interfere constructively or destructively depending on the choice of horizontal and diagonal tunneling terms ($t_h$ and $t_d$, respectively). 
    }
\label{fig:LGT}
\end{figure}

At its core, particle physics forms the foundation of our comprehension of the fundamental workings of the universe, explaining how matter and forces interact via gauge degrees of freedom. Quantum simulators are a natural tool to simulate fundamental interaction beyond classical capabilities \cite{Ba_uls_2020}. To simulate the interaction of fundamental forces, Lattice Gauge Theories (LGT) describe models where a matter degree of freedom is coupled to a gauge field. For example, a Hamiltonian defined on a lattice where the hopping of the matter from one site to the next one is mediated by a gauge field. In this scenario, it is thus convenient to use the synthetic dimensions as a degree of freedom where either the gauge or the matter field can be encoded, and many proposals have been formulated to pursue this approach \cite{Halimeh:2023lid}. 

 The Schwinger model represented in Fig.~\ref{fig:LGT}(a) is an illustrative example of the mapping between the matter and gauge degrees of freedom to internal states of an atomic species in an optical lattice~\cite{suraceInitio2023a}. It describes the interactions between massive fermionic particles, $\hat \psi_j$, living in site $j$ of a 1D lattice (matter degree of freedom), which are influenced by a U(1) gauge field described by the electric field, $\hat U_{j,j+1}$, placed at the bonds. Despite the apparent simplicity of this 1+1 dimensional model of quantum electrodynamcis, it has a non-perturbatively generated mass gap and shares some features with Quantum Chromodynamics (QCD), such as confinement and chiral symmetry breaking, and has been adopted as a benchmark model where to explore LGT techniques.

As an extension of this minimal model, Ref.~\citeonline{tagliacozzoOptical2013} studies invariant LGTs in 2+1 dimensions, where one observes a spontaneous breaking of the gauge symmetry, as well as charge confinement. Different experimental platforms have addressed these simplified models. For example, long-range Rydberg interactions can be used to ensure the Gauss law of the theory\cite{suraceLattice2020}, or help with the simulation of non-abelian theories~\cite{tagliacozzoSimulation2013}. In the latter case, the synthetic dimension is not only in the internal state, but also on where the excitation is in the superlattice. 
  Experimentally, a simplified bosonic version with $Z_2$ gauge fields has also been  implemented using two atomic species in a one-dimensional optical lattice~\cite{schweizerFloquet2019}. Following Sec.~\ref{Sec3}, another possibility consists in realizing exotic geometries through Raman-assisted tunnelings\cite{suszalskiDifferent2016}, and Ref.~\citeonline{tagliacozzoTensor2014} further shows how to encode the symmetries in a tensor-network architecture.

LGTs can also offer deep connections with topology, as it can be illustrated with the Creutz-ladder model~\cite{creutzEnd1999} depicted in Fig.~\ref{fig:LGT}(b). There, fermionic atoms are trapped in a 1D optical lattice and two internal atomic states, $\downarrow,\uparrow$, are seen as an orthogonal spatial degree of freedom. This results into an effective two-leg ladder Hamiltonian with crossed links where atoms can tunnel along the horizontal ($t_h$), vertical ($t_v$) an diagonal directions ($t_d$). Using assisted tunneling, one can obtain complex coefficients of the form $\tilde t_h=e^{\pm i\theta }t_h$, which translate into a net flux $2\theta$ when a fermion hops around a square unit cell. 
Interestingly, for a configuration with $\theta=\pi/2$ and null vertical tunneling, $t_v=0$, $t_h=t_d$, the two possible paths an electron can follow to move two sites (depicted in yellow and blue) interfere destructively. In the dispersion relation for noninteracting particles, this manifests as gapped flat bands with an associated vanishing group velocity. Therefore, if any transport occurs in a system with such a geometric frustration, that must be related to interactions among the fermionic particles. Furthermore, these flat bands can also sustain many nontrivial effects, including topological and chiral states.
Different experimental schemes have been proposed to engineer the needed effective tunnelings of the Creutz-ladder model using 1D atomic chains and one additional synthetic dimension~\cite{tovmasyanGeometryinduced2013}, including Raman-assisted tunneling~\cite{junemannExploring2017} or Floquet engineering through shaken lattices~\cite{gorgRealization2019,eckardtColloquium2017}.
Focusing on LGTs, the interplay between spinless fermions interacting with a dynamical $\mathds{Z}_2$ gauge fields in a Creutz-ladder geometry also represents a minimal model for gauge theories, manifesting characteristic features such as deconfinement and topological order~\cite{gonzalez-cuadraRobust2020}.

\begin{figure}[t!]
\centering
\includegraphics[width=14cm]{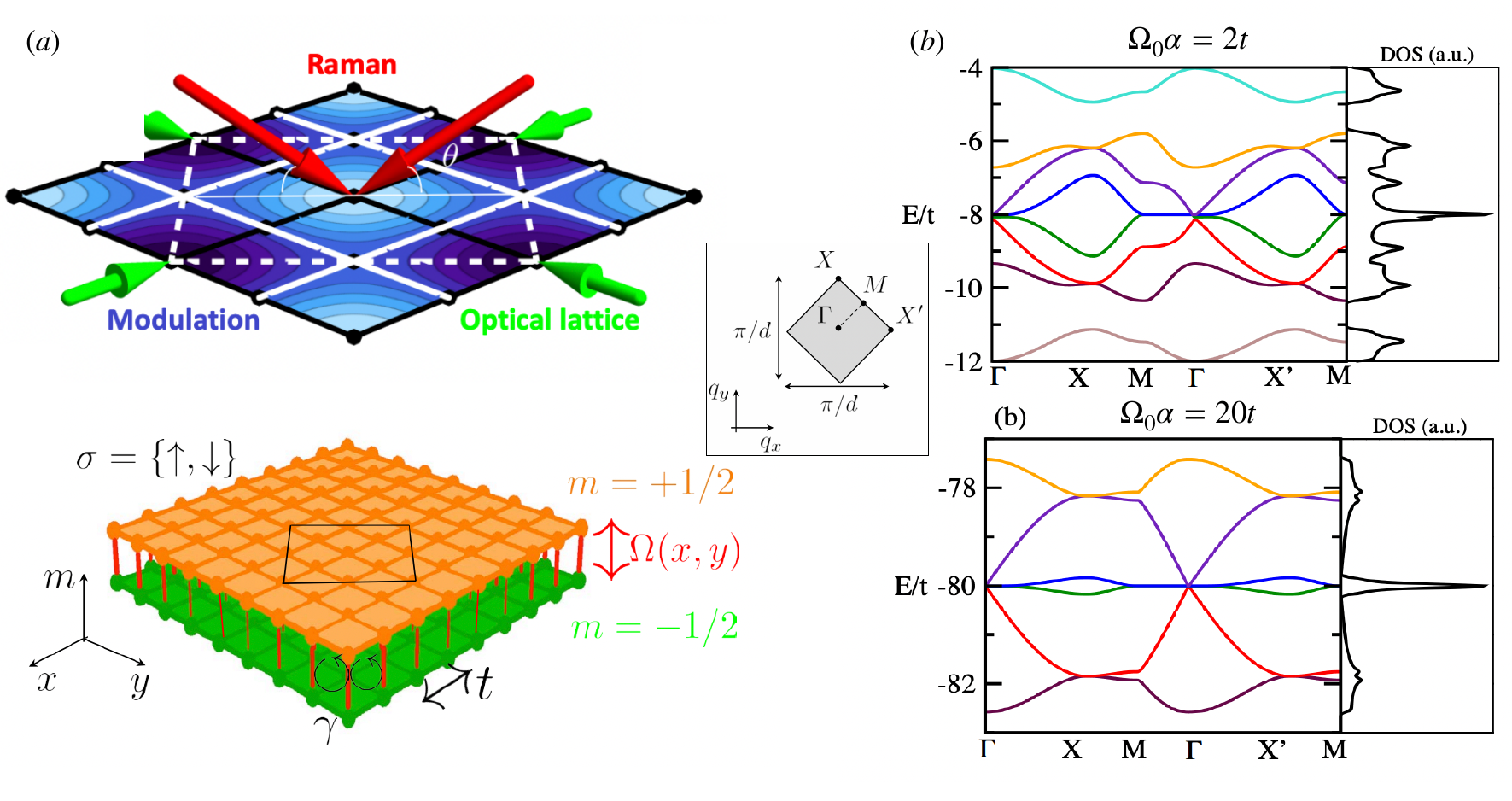}
\caption{ Twistronics without a twist: the case of a synthetic bilayer. \textbf{Panel (a,top)}: 
A suitable two-dimensional Fermi gas (e.g. $^{87}$Sr) with four distinct magnetic sub-levels chosen from the ground state manifold labelled by a pair of two-valued quantum numbers $\{\sigma, m\}$  is trapped in a single layer state independent optical lattice,  chosen here as a square lattice.   This system forms a synthetic bilayer if  one of the  quantum numbers, say,  $m=\pm 1/2$ is identified with the layer degree of freedom.  The
fermion spin degree of freedom within a synthetic layer is given by $\sigma =\uparrow,\downarrow$.  Each fermion species can tunnel between sites of the optical lattice with hopping parameter $t$.  Additional Raman coupling $\Omega_0$ can be utilized to induce transitions between $m=+1/2$ and $m=-1/2$ states effectively introducing (in general,  tunable complex valued) interlayer hopping between the synthetic layers.  An appropriate scheme utilzing  a spatial light modulator can be used to engineer spatially modulated Raman coupling $\Omega(x,y)$  leading to  systems with  Moir\'e unit cell patterns. \textbf{(Panel (a), bottom)}  shows the synthetic bilayer obtained with $\Omega(x,y) = \Omega_0 \left[1 -\alpha(1+\cos{(2 \pi x/l_x)} \cos{(2 \pi y/l_y)}) \right]$, where $l_x$ and $l_y$ represent the periodicities along the $x$ and $y$ axes, respectively.  \textbf{(Panel (b))} Tunable quasi-flat bands and Dirac cone  spectra appear for special choices of periodicities. Shown here are bandstructures for $(l_x,l_y)=(4,4)$.  Upper plot represents the negative part of the spectrum along the high symmetry points (we omit the postive part which is symmetric with respect to $E=0$) for $\Omega_0\alpha=2t$, while lower plot depicts strong bands flattening at $\Omega_0\alpha=20t$. Figure adapted from Ref.~\citeonline{Salamon_20}.}
\label{fig:twistronics} 
\end{figure}

\section{Twistronics \label{Sec5}}

In recent years,  Moir\'e 
materials have emerged as a new platform for strongly correlated phenomena.  These materials consist of stacked layers coupled  via van der Waals forces where periodicity mismatch or twisting (i.e.  rotational misalignment) between layers leads to long wavelength Moir\'e patterns in real space resulting under certain conditions in significant renormalization of parent bandstructures. In the paradigmatic example of graphene bilayers, twisting was theoretically predicted to lead to a strong reduction of the Fermi velocity~\cite{CastroNeto2007,Barticevic2010} resulting in almost flat bands  at low energies at special so-called magic angles~\cite{Bistritzer2010}.  The kinetic energy scales are thus effectively quenched and interaction effects can become dominant and support  the  emergence of new correlated ground states at partial filling.  These expectations were spectacularly confirmed experimentally in magic angle twisted bilayer graphene revealing superconducting domes,  correlated insulating states as well as strange metallic behaviour - properties absent in the underlying monolayers - triggering  the intense new research field of  twist induced electronic phenomena (or twistronics)  and more generally Moir\'e superlattice structures~\cite{Andrei2021,Kennes2021}.

The novel control over material properties via Moir\'e pattern engineering  comes with certain challenges. In the context of bilayer graphene, for instance, very small magic angles of the order of  $\theta \sim 1^\circ$  were predicted. Such small angles in layered systems are experimentally challenging to stabilize and maintain homogeneously over extended spatial regions. Moreover, fundamentally, they lead to very large unit cells containing thousands of monolayer atoms each rendering microscopic first principles studies and theoretical understanding of the emerging physics difficult. Such considerations have strongly motivated the theoretical design of highly controllable cold atom based quantum simulators of twistronic or Moir\'e materials.

As already mentioned, cold atoms trapped in optical lattices are  highy versatile platforms for the realization of a plethora of models of interest in condensed matter physics,  including graphene like systems~\cite{Tarruell2012}.  
The possibilities of creating various lattice geometries,   and the choice of particle statistics~\cite{Esslinger_2010} contribute to the variety of many-body problems that can be simulated with cold atoms  (see \textit{e.g.}  Ref.~\citeonline{Grass2016} for a proposal of coupling a graphene-like layer to a non-matching square lattice substrate).  Most importantly,   the physical parameters of such lattice systems can be  tuned well beyond regimes attainable in the solid state. 
In particular,  in the context of Moir\'e physics,  the interlayer coupling strength can be crucially tuned  to high values which can lead to larger magic angles
and consequently small Moir\'e unit cells.  
This feature may in principle simplify theoretical modelling and help unveil  the mechanisms behind various physical properties of Moiré systems.

The mapping of two (or more) long lived internal degrees of trapped cold atoms  into  the  layer degree of freedom  is ideally suited for  quantum simulation of  twistronics or Moir\'e physics.  
The basic challenge here is how to implement the twist using the idea of synthetic dimensions.  Two different approaches have been put forward.  
In the spirit of physical twisting in materials,  a scheme  where two internal states  are subjected independently  to two state dependent optical lattice potentials rotated by an angle with respect to each other was introduced in Ref.~\citeonline{Gonz_lez_Tudela_2019}.   The respective atomic excitations can then hop within each lattice layer,  while effective interlayer hopping is induced and controlled via Raman coupling of the two internal states.   
Magic angle phenomena were predicted for such synthetic systems upto twist angle $\theta \sim 6^{\rm o}$ \cite{Gonz_lez_Tudela_2019,Luo_2021}.   The first proof-of-principle experimental realization of this twisted synthetic system has been recently achieved  \cite{Meng2023} with bosonic atoms, demonstrating (i) the emergence of a Moir\'e supercell,  and (ii) the tunability of the effective interlayer coupling.  
   
A different scheme for simulating twistronics intriguingly without a physical twist was introduced in Ref.~\citeonline{Salamon_20} (see Fig.\ref{fig:twistronics}). 
This scheme builds on the idea that physical twisting of two layers fundamentally leads to,  and therefore can be mimicked by, spatial modulation of interlayer coupling on a lattice  (note also alternate proposals focussing on inducing magic angle phenomena in layered systems with quasiperiodic potentials ~\cite{Fu2020,Chou2020,Fu2021}).    
Considering   two internal  states  (labelling two synthetic layers) of an appropriately chosen atomic species trapped by a single two-dimensional optical lattice potential,  
it was shown that such spatially dependent interlayer hoppings can be directly imprinted on the lattice via specifically designed spatial control of Raman coupling ~\cite{Salamon_20}. 
While this scheme is quite general,  it in particular allows the creation of Moir\'e systems with a small unit cell where the control of the strength, phase and  periodicity of the Raman coupling 
leads to a broad range of band structures including quasi-flat bands with tunable widths for special magic values of periodicity.   
The periodic modulation shown in Fig \ref{fig:twistronics} supports e.g. topological effects such as the anomalous quantum Hall effect via the control of hopping parameters~\cite{Salamon_20_2} as well as interaction effects such as flatband superconductivity~\cite{Salamon_2022}.  

The theoretical and experimental results so far foreshadow many possibilities of cold atom based twistronics.  On one hand, it is  natural to foresee that synthentic dimensions can be used to simulate mutlilayer twistronics  where more than two internal layers are utlized and offer the intriguing possiblity of observing and uncovering interaction and topological effects for boson, fermion and spin systems.  From a fundamental perspective,  the absence   of electron-phonon coupling in cold atom systems could \textit{e.g.} shed light on the relative importance of such coupling in  phenomena observed in Moir\'e materials. 

\section{Quantum random walks\label{Sec6}}

Quantum walks (QWs) are deterministic quantum counterparts to classical random walks, where a particle (quantum walker) performs discrete steps conditioned by the instantaneous configuration of its spin-like degree of freedom. A walker with a binary ``spin'' degree of freedom engages in a series of unitary operations to determine its movement between neighboring lattice sites. The quantum evolution is realized through the repeated application of a unitary operation that defines each step the walker performs.  
Quantum walks provide a versatile platform for study dynamics within a wide spectrum of topological phases, and it has been demonstrated that QWs can encompass all feasible symmetry-protected topological phases observed in non-interacting fermions within one or two spatial dimensions (1D or 2D) \cite{PhysRevA.82.033429,PhysRevA.94.023631}. 

In \cite{Cardano2017,Maffei_2018}, the authors proposed and experimentally validated a method for discerning topological characteristics within the bulk of one-dimensional chiral systems via the introduced concept of the mean chiral displacement, an observable that rapidly converges to a value proportional to the Zak phase during the system's free evolution. The measurement of the Zak phase in a photonic quantum walk employing twisted photons is achieved by observing the mean chiral displacement (MCD) within its bulk. The MCD is a potent tool for probing the topology of chiral $1D$ systems whose initial state is connected to a localized state through a unitary and translation-invariant transformation. Consequently, MCD serves as a topological indicator in experiments involving abrupt transitions between different topological phases in the study of topological systems undergoing dynamic phase changes and out-of-equilibrium dynamics \cite{PhysRevResearch.2.023119,DiColandrea_2022,DiColandrea:23}

Finally,  photonic simulation of a two-dimensional quantum walk\cite{DErrico:20} was proposed. In this scenario, the positions of the walkers are encoded in the transverse wavevector components of a single light beam. The desired dynamics is achieved through a sequence of liquid-crystal devices, which impart polarization-dependent transverse kicks to the photons in the beam. This engineered quantum walk realizes a periodically-driven Chern insulator, and its topological features are probed by detecting the anomalous displacement of the photonic wavepacket under the influence of a constant force. This compact and versatile platform offers promising opportunities for simulating two-dimensional quantum dynamics and topological systems.

\section{Time Crystal Platform for Quantum Simulations\label{Sec7}}

During the space crystal formation, the continuous space translational symmetry is spontaneously broken due to the many-body interactions and a regular distribution of atoms emerges. As such, space crystals are characterized by a discrete space translation symmetry. The condensed matter crystalline structures reveal many different phases of matter ranging from band and Mott insulators to topological phases, which can be studied in \textit{standard} quantum simulators. Recently, a new paradigm appeared to simulate exotic phases of matter in the \textit{time domain}\cite{Sacha2015,PhysRevA.94.023633,PhysRevA.97.053621,PhysRevLett.120.140401}, allowing studies of condensed matter physics in time crystalline structures - the novel platform for quantum simulations.

The idea of quantum time crystals was introduced by Wilczek in 2012, Ref. \citeonline{PhysRevLett.109.160401}. However, it was later proven that for a wide class of systems with a two-body finite range of interactions, spontaneous breaking of time translational symmetry, i.e., the formation of time crystals, is impossible in the ground state \cite{PhysRevLett.111.070402, PhysRevLett.114.251603, Watanabe2020, PhysRevLett.119.250602}. Nevertheless, his idea led to the discovery of discrete time crystals (DTC) and the beginning of solid-state physics in the time domain\cite{PhysRevA.91.033617,  PhysRevLett.116.250401, PhysRevLett.117.090402, Zhang2017, Choi2017}.

DTCs are time-periodic phases of matter that spontaneously break the discrete time-translation symmetry $t \to t + T$ down to $t \to t + nT$ for some integer $n > 1$ \cite{PhysRevA.91.033617,PhysRevLett.116.250401,PhysRevLett.117.090402, PhysRevB.94.085112, PhysRevLett.118.030401, Kuros_2020}. Research on time crystals led to the creation of a new platform of quantum simulators, where time can be used, analogous to artificial dimensions, to study multidimensional structures (for an extensive introduction to the field, we refer to  \cite{Sacha_2018,Sacha2020, Else2020, Guo_2020, RevModPhys.95.031001}).
The time-crystalline structures open a new research direction in the field of quantum simulators of topological matter \cite{Giergiel_2019}, allowing simulation of paradigmatic topological models like the Su-Schrieffer-Heeger model or Bose-Haldane model, realized in the time domain, with the bulk-edge correspondence related to the edge localized in time.
In particular, the quasienergy spectrum of a resonantl driven optical lattice may be interpreted as that of a crystal-like structure
with the time playing the role of an additional coordinate\cite{Braver2022}. With this analogy, authors studied adiabatic variation
of the driving protocol and demonstrated that it leads
to a change of system dynamics that is a manifestation
of the Thouless pumping in the temporal dimension. 
Moreover, topological effects emerging due to nontrivial time lattice geometry have been studied in Ref.\citeonline{PhysRevLett.127.263003}, where the authors showed that inseparable two-dimensional time lattices with the M\"obius strip geometry could be realized for ultracold atoms bouncing between two periodically oscillating mirrors, a Lieb lattice model with a flat band can be realized. 

 Time crystals also offer a platform for simulating higher-dimensional topological models in the time domain via periodically ordered physical structures, where time is the additional coordinate. The time-crystalline approach involves a driving signal of a certain frequency to create a repeating pattern of motion at a commensurate frequency that persists over time. Many condensed matter phenomena were thus reenacted in the time domain, and the possibility to engage both temporal and spatial dimensions at the same time was established, thus doubling the number of available dimensions.  Combining time and space crystalline structures makes it possible to realize a $6$-dimensional time-space crystals for a resonantly driven $3$-dimensional \cite{Zlabys2021}, allowing observation of a $6$-dimensional quantum Hall effect.
Next, the proposal for a $8$-dimensional system\cite{PhysRevB.108.L020303}  utilizes only two physical spatial dimensions.  The topological nature of the attained time-space crystalline structure is evident by considering adiabatic state pumping along temporal and spatial crystalline directions. Interpreting the two adiabatic phases as crystal momenta of simulated extra dimensions, authors showed that non-vanishing second Chern numbers of the effective $4$-dimensional lattice characterize the energy bands of the system. The $N$-dimensional crystalline structure simulator can be realized considering the system of $N$-bouncing particles on an oscillating mirror. For a specific mirror oscillation frequency, the system can behave like an $N$-dimensional fictitious particle moving in an $N$-dimensional crystalline structure \cite{Golletz_2022}.
 
\section{Experimental perspectives\label{Sec8}}

Since the first original proposals \cite{Boada_2011, celiSynthetic2014}, synthetic dimensions have been experimentally realized using a broad range of degrees of freedom. In atomic systems, these include sublevels of the atomic ground state (with experiments exploiting rubidium \cite{Stuhl_2015}, ytterbium \cite{Mancini_2015} and dysprosium \cite{ChalopinNaturePhys2020} atoms), ground and metastable ``clock'' states \cite{LiviPRL2016, KolkowitzNature2017}, momentum states \cite{MeierPRA2016, ChenNJPQI2021}, Rydberg states \cite{KanungoNatCommun2022, ChenArXiv2023}, and harmonic trap states \cite{OliverPRR2023}. Moreover, synthetic dimensions have also been implemented in photonic systems, exploiting angular momentum modes \cite{Cardano2017} and time bin modes \cite{ChalabiPRL2019}. 

Focusing on the atomic platforms, and more specifically on the use of atomic internal states, the synthetic dimension approach has provided access to new classes of experiments that were not possible using conventional real-space realizations. Following the original proposal discussed in Sec. \ref{Sec3}, a prime example is the implementation of strong synthetic magnetic fields for the atoms, both in the lattice (realizing the celebrated Hofstadter model) \cite{Stuhl_2015, Mancini_2015} and in the continuum \cite{ChalopinNaturePhys2020}. In both cases, a key advantage is that the system has sharp boundaries along the synthetic dimension, enabling the direct visualization of the skipping orbits associated to the topologically protected chiral edge states. Moreover, by coupling the synthetic dimensions in a cyclic manner, systems with periodic boundary conditions can be engineered. This has led to the realization of synthetic Hall cylinders \cite{HanPRL2019, FabrePRL2022}, where Laughlin's topological charge pump thought experiment was recently investigated experimentally, proving the nontrivial topology of quantum Hall insulators. Another exciting research direction, which closely follows the original motivation of the synthetic dimension approach (see Sec. \ref{Sec1}), is the realization of atomic systems in more than three spatial dimensions. Extra dimensions are encoded in the additional degrees of freedom, enabling for instance the investigation of four-dimensional quantum Hall systems \cite{BouhironArxiv2022}. 

All the experiments discussed above focus on non-interacting physics. However, in atomic systems synthetic dimensions offer a promising route towards the realization of strongly-correlated systems. On the one hand, the energy splitting of the internal atomic levels can naturally be made much larger than all the other energy scales of the system, thus avoiding the heating problems associated to Floquet engineering approaches often used to implement artificial gauge fields \cite{WeitenbergNatPhys2021}. On the other hand, the optical power required to optically couple the atomic internal states remains low, keeping the heating associated with inelastic photon scattering to acceptable levels. Hence, exploiting synthetic dimensions to investigate the rich many-body physics of quantum Hall systems seems within experimental reach, as recent measurements of the Hall response of strongly interacting synthetic ytterbium ladders demonstrate \cite{ZhouScience2023}.

The investigation of strongly-interacting systems using synthetic atomic dimensions will certainly experience a rapid development in the coming years. Promising research directions are the investigation of the many-body phase diagram of interacting Hofstadter ladder systems, where the ground state and quench dynamics of the system seem within experimental reach \cite{BuserPRA2020} and even magnetic frustration in an effective triangular geometry could be investigated \cite{BarbieroPRR2023}. In synthetic dimensional systems, the interactions along the synthetic dimension acquire a peculiar long-range character. While this is normally seen as a nuisance, leading to the development of schemes to cancel them \cite{Barbiero_2020}, it could also by leveraged as a resource. For instance, building on previous work in the continuum \cite{FroelianNature2022, ChisholmPRR2022}, synthetic dimension interactions could be exploited to realize one-dimensional anyon Hubbard models. Another advantage of synthetic dimensions is that the tunneling along the synthetic dimension can be made spatially periodic. As discussed in Sec. \ref{Sec5}, this idea could be exploited to engineer bilayer systems with tunable supercells, providing  access to twistronic physics in atomic physics platforms \cite{Salamon_2022}. Finally, by extending the control over the synthetic tunneling to each individual synthetic site, completely new classes of experiments become possible, such as the engineering of the entanglement Hamiltonian of a quantum Hall system \cite{RedonArxiv2023}, or the simulation of infinite size many-body systems in finite size quantum simulators \cite{KuzminPRXQ2022}. 

\section{Conclusions\label{Sec9}}

This Perspective article comprehensively illustrates the profound significance of synthetic dimensions in contemporary quantum research. We have summarized the remarkable potential of employing synthetic dimensions in quantum simulators, enabling the faithful simulation of exotic space-time phenomena and facilitating the investigation of artificial gauge fields and lattice gauge theories within the ultra-cold quantum gases in optical lattices. Furthermore, synthetic dimensions permit the emulation of ``twistronics without a twist,'' offering a novel approach to research of strongly correlated materials. Moreover, the employment of quantum random walks within the synthetic dimensions framework opens avenues for studying a broad spectrum of symmetry-protected topological phases, as observed in photonic systems in one or two spatial dimensions. Finally, discrete time crystals have emerged as a groundbreaking platform for quantum simulation. Discrete time crystals enable the investigation of high-dimensional topological models by introducing time as an additional artificial dimension. This discourse underscores the noteworthy progress made in recent years in the experimental realization and application of synthetic dimensions, thereby underlying its growing prominence and relevance within the quantum research landscape.

\section*{Acknowledgements}
ICFO QOT group acknowledges support from:
ERC AdG NOQIA; MCIN/AEI (PGC2018-0910.13039/501100011033, CEX2019-000910-S/10.13039/501100011033, Plan National FIDEUA PID2019-106901GB-I00, Plan National STAMEENA PID2022-139099NB-I00 project funded by MCIN/AEI/10.13039/501100011033 and by the “European Union NextGenerationEU/PRTR'' (PRTR-C17.I1), FPI); QUANTERA MAQS PCI2019-111828-2); QUANTERA DYNAMITE PCI2022-132919 (QuantERA II Programme co-funded by European Union’s Horizon 2020 program under Grant Agreement No 101017733), Ministry of Economic Affairs and Digital Transformation of the Spanish Government through the QUANTUM ENIA project call – Quantum Spain project, and by the European Union through the Recovery, Transformation, and Resilience Plan – NextGenerationEU within the framework of the Digital Spain 2026 Agenda; Fundació Cellex; Fundació Mir-Puig; Generalitat de Catalunya (European Social Fund FEDER and CERCA program, AGAUR Grant No. 2021 SGR 01452, QuantumCAT \ U16-011424, co-funded by ERDF Operational Program of Catalonia 2014-2020); Barcelona Supercomputing Center MareNostrum (FI-2023-1-0013); EU Quantum Flagship (PASQuanS2.1, 101113690); EU Horizon 2020 FET-OPEN OPTOlogic (Grant No 899794); EU Horizon Europe Program (Grant Agreement 101080086 — NeQST), ICFO Internal “QuantumGaudi” project; European Union’s Horizon 2020 program under the Marie Sklodowska-Curie grant agreement No 847648; “La Caixa” Junior Leaders fellowships, La Caixa” Foundation (ID 100010434): CF/BQ/PR23/11980043.
RWC acknowledges support from the Polish National Science Centre (NCN) under the Maestro Grant No. DEC- 2019/34/A/ST2/00081.
M.P. acknowledges the support of the Polish National Agency for Academic Exchange, the Bekker programme no:
PPN/BEK/2020/1/00317.
A.C. acknowledges the support of MCIN/AEI/10.13039/501100011033 (LIGAS PID2020-112687GB-C22) and Generalitat de Catalunya (AGAUR 2021 SGR 00138).
L. T. acknowledges funding from the European Union (ERC CoG-101003295 SuperComp), MCIN/AEI/10.13039/501100011033 (LIGAS project PID2020-112687GB-C21, DYNAMITE QuantERA project PCI2022-132919 with funding from European Union NextGenerationEU, Severo Ochoa CEX2019-000910-S, and PRTR-C17.I1 with funding from European Union NextGenerationEU and Generalitat de Catalunya), Deutsche Forschungsgemeinschaft (Research Unit FOR2414, Project No. 277974659), Fundación Ramón Areces (project CODEC), Fundació Cellex, Fundació Mir-Puig, and Generalitat de Catalunya (ERDF Operational Program of Catalunya, Project QUASI-CAT/QuantumCat Ref. No. 001-P-001644, CERCA program, and AGAUR 2021 SGR 01448).
Views and opinions expressed are, however, those of the author(s) only and do not necessarily reflect those of the European Union, European Commission, European Climate, Infrastructure and Environment Executive Agency (CINEA), or any other granting authority. Neither the European Union nor any granting authority can be held responsible for them.
T.G. acknowledges funding from
BBVA Foundation (Beca Leonardo a Investigadores en Física 2023), Gipuzkoa Provincial Council (QUAN-000021-01), the Department of Education of the Basque Government (PIBA\_2023\_1\_0021 (TENINT)). The BBVA Foundation is not responsible for the opinions, comments and contents included in the project and/or the results derived therefrom, which are the total and absolute responsibility of the authors.

\section*{Author contributions}
The authors contributed equally to all aspects of the article.  

\section*{Competing interests}
The authors declare no competing interests.  

\section*{Publisher’s note}
Springer Nature remains neutral with regard to jurisdictional claims in published maps and institutional affiliations.

\section*{Supplementary information (optional)}
None.

\bibliography{bibl}

\end{document}